\documentclass[twocolumn,preprintnumbers,amsmath,amssymb,floatfix,showpacs,prl]{revtex4}

\usepackage{dcolumn}
\usepackage{bm}
\usepackage{graphicx}

\begin{document}


\title{Self-duality and long-wavelength behavior of the Landau-level guiding-center
structure function, and the shear modulus of fractional quantum Hall fluids}

\author{F. D. M. Haldane}
\affiliation{Department of Physics, Princeton University,
Princeton NJ 08544-0708}

\date{December 6, 2011}
\begin{abstract}
A remarkable self-duality of the ``guiding-center'' structure function
$s(\bm q)$ of  particles in a partially-filled 2D Landau level is
used to express the long-wavelength collective
excitation energy of fractional quantum Hall fluids in terms of a shear
modulus. A bound on the small-$\bm q$ behavior of $s(\bm
q)$ is given.
\end{abstract}
\maketitle

Landau quantization of the kinetic energy of charged particles moving
on a two-dimensional surface through which a uniform magnetic flux
density $B$ passes
leaves their non-commuting ``guiding-center'' coordinates as the
residual degrees of freedom.    The  (static) ``guiding-center structure
factor'' $\bar s(\bm q)$  introduced by Girvin, MacDonald, and
Platzman (GMP)\cite{GMP} characterizes the two-particle correlation function and plays
a major role in their ``single-mode approximation'' (SMA) treatment 
of collective excitations of
the Laughlin incompressible fractional quantum Hall (FQH)
state\cite{laughlin}.      

 The problem of interacting guiding
centers is essentially a \textit{quantum geometry} problem.    I
will show that the guiding-center structure factor has a remarkable self-duality that (for single-component systems
with no other internal degrees of freedom, such as spin-polarized
electrons or spinless bosons) makes it proportional (after a rotation)
to its own Fourier transform.

Using an analogy to Feynman's treatment\cite{feynman} of the
collective mode in  superfluid  $^4$He,
GMP found an upper bound to the Laughlin-state collective-mode
dispersion in the form $\bar f(\bm q)/\bar s(\bm q)$, where
$\lim_{\lambda\rightarrow 0}\bar f(\lambda \bm q)$ $\propto$ $\lambda^4$.
Using the self-duality to simplify the
GMP expression for $\bar f(\bm q)$, I will relate its
long-wavelength limit 
to a ``shear modulus'' of the incompressible FQH fluid,
which describes the energy cost of distorting the shape of the
``elementary droplet''  of the FQH fluid.
This shape can be viewed as the shape of the ``attached flux''
of the ``composite boson'' of the fluid, and plays a central role
in a ``geometric field theory'' description of the FQH\cite{haldane2011}.

In an  idealized model,  charge-$e$ particles move on a
flat translationally-invariant 2D surface. The
``magnetic area'' through which a London flux quantum $\Phi_0$ = $h/e$ 
passes is $2\pi \ell_B^2$.   A particle guiding-center $\bm R$ =
$R^a\bm e_a$,
where $\{\bm e_a, a=1,2\}$
are   orthonormal tangent vectors of  the surface,
has non-commuting components
\begin{equation}
[R^a,R^b] = -i\epsilon^{ab}\ell_B^2,
\label{cr}
\end{equation}
where $\epsilon^{ab}$ is the antisymmetric (2D Levi-Civita) symbol.

In order to make any (non-topological) metric-dependence explicit,
I use a (spatially) covariant notation that distinguishes
displacements (with upper indices) from (reciprocal) vectors (with
lower indices), with Einstein summation convention over repeated
upper/lower index
pairs.
The tangent vectors are chosen orthonormal ($\bm e_a\cdot \bm e_b$ =
$\eta_{ab}$)  with respect to the 2D Euclidean metric $\eta_{ab}$,
induced by the 3D  Euclidean metric of the space in which the
surface is embedded. This metric is important at atomic scales, but
plays no fundamental role in FQH systems where the  quantum-geometric 
``magnetic area'' $2\pi\ell_B^2$
is much larger than the area of any atomic-scale unit cell on the
surface, and provides  the ``ultraviolet regularization''
 of the continuum description.

If the energy splitting of the  degenerate Landau levels dominates the interaction
energies, and  a single Landau level  is partially occupied,  the
leading term in the residual effective Hamiltonian that  lifts the
degeneracy of the partially-filled Landau level  by two-body interactions is
\begin{eqnarray}
&& H = \int \frac{d^2\bm q\ell_B^2}{4\pi} \tilde v(\bm
q)\rho(\bm q) \rho(-\bm q) 
\label{ham}
\\
&& \rho(\bm q) = \sum_{\alpha\alpha'}\langle \alpha |e^{i\bm
  q\cdot \bm R}|\alpha'\rangle 
  c^{\dagger}_{\alpha}c_{\alpha'},
\end{eqnarray}
where $c^{\dagger}_{\alpha}$ creates a particle in state
$|\alpha\rangle$, and $\alpha$ labels orbitals in an arbitrary
orthonormal basis of  states of the Landau level.  
Here $\rho (\bm q)$ obeys the Lie algebra
\begin{equation}
[\rho(\bm q), \rho(\bm q') ] = 2i\sin ({\textstyle\frac{1}{2}}\bm
q\times \bm q'\ell_B^2)\rho(\bm q + \bm q'),
\label{gmpa}
\end{equation}
where $\bm q \times \bm q'$ $\equiv$  $\epsilon^{ab}q_aq'_b$, $q_a $ =
$\bm q \cdot \bm e_a $.
The fundamental algebra (\ref{gmpa})
was first noticed in this context by GMP\cite{GMP}.  
The operators $\rho(\bm q)$ with $\bm q \ne 0$ are 
generators of  area-preserving diffeomorphisms (APD) of the Landau
level.
A filled Landau-level  is
invariant under a APD, 
but  FQH states are not\cite{suss}.

The nature of the ground state of (\ref{ham}) depends on the (real)
interaction  $\tilde v(\bm q)$ = $\tilde v(-\bm q)$, which is a product of the Fourier transform
of the Coulomb interaction potential between the particles, a form
factor of the quantum well which binds them to the 2D surface, and a
 Landau orbit form-factor.   The form factors ensure that $v(\bm r)$, where
\begin{equation}
H = \sum_{i<j}v(\bm R_i - \bm R_j), \quad 
v(\bm r) = \int \frac{d^2\bm q\ell_B^2}{2\pi} \tilde v(\bm q)  e^{i\bm q\cdot
  \bm r}, 
\end{equation}
is finite.
For general $\tilde v(\bm q)$, the Hamiltonian (\ref{ham}),  has both translational
and 2D inversion symmetry ($\bm R \rightarrow -\bm R$).
The interaction also  has  rotational invariance if
\begin{equation}
\tilde v(\bm q) = \tilde v(q_g), \quad  q_g^2 \equiv g^{ab}q_aq_b,
\end{equation} 
where $g^{ab}$ is the inverse of a positive-definite unimodular
(determinant =  1) metric $g_{ab}$; this will only occur if
the shape of the Landau orbits are congruent with the shape of the
Coulomb equipotentials around a point charge on the surface.  
In
practice, this only happens when there is an atomic-scale three-fold
or four-fold rotation axis normal to the surface, and no ``tilting'' of
the magnetic field relative to this axis, in which case $g_{ab}$ =
$\eta_{ab}$.

I will  assume that translational symmetry is unbroken, so
$\langle c^{\dagger}_{\alpha}c_{\alpha'}\rangle$ = $\nu \delta_{\alpha\alpha'}$,
where $\nu$ is the ``filling factor'' of the Landau level.  
(In a 2D system, this will
always be true at  finite temperatures, but may break down as
$T\rightarrow 0$).
Then $\langle \rho(\bm q)\rangle$ = $2\pi \nu \delta^2(\bm q\ell_B)$.
Note that the fluctuation  $\delta \rho(\bm q)$ = $\rho(\bm q) - \langle \rho(\bm
q)\rangle$
also obeys the algebra (\ref{gmpa}).
I will define a guiding-center structure factor  $s(\bm q)$ = $s(-\bm
q)$ by
\begin{equation}
{\textstyle\frac{1}{2}}\langle \{\delta\rho (\bm q),\delta \rho(\bm q')\}
  \rangle = 2\pi  s(\bm q)\delta^2(\bm q\ell_B + \bm q'\ell_B)
.
\end{equation}
This is a structure factor defined per flux quantum, and is given 
in terms of the  GMP structure factor  $\bar s(\bm q)$ of Ref.\cite{GMP}  (defined
per particle) by $s(\bm q)$ = $\nu\bar s(\bm q)$.
I also define $s^a(\bm q)$ $\equiv$
$\partial s(\bm q)/\partial q_a$, $s^{ab}(\bm q)$ $\equiv$
$\partial^2s(\bm q)/\partial q_a\partial q_b$, \textit{etc.}

 In the ``high-temperature limit'' where
$|v(\bm r) | \ll k_BT$ for all $\bm r$, but $k_BT$ remains much smaller than the gap between
Landau levels, the guiding centers become completely uncorrelated,
with $\langle
c^{\dagger}_{\alpha}c_{\alpha'}
c^{\dagger}_{\beta}c_{\beta'}\rangle$
$-$ $\langle
c^{\dagger}_{\alpha}c_{\alpha'} \rangle
\langle c^{\dagger}_{\beta}c_{\beta'}\rangle$
$\rightarrow$ 
$s_{\infty}\delta_{\alpha\beta'}\delta_{\beta\alpha'}$,
with
$s_{\infty}$ = $\nu + \xi \nu^2$,
where $\xi$ = $-1$ if the particles are spin-polarized fermions, and
$\xi $ = $+1$ if they are bosons (which may be relevant for cold-atom systems).
Note that for all 
temperatures,  
$\lim_{\lambda \rightarrow \infty} s(\lambda \bm q)$ = $s_{\infty}$, while
$s(0)$ = $\lim_{\lambda \rightarrow 0} s(\lambda \bm q)$ = 
$k_BT/\left (\left .\partial^2f(T,\nu)/\partial \nu^2\right |_T\right )$, 
where $f(T,\nu)$ is the free energy per flux quantum.
$s(0)$ vanishes at $T$ = 0, and at all $T$ if $\tilde v(\lambda \bm q)$ diverges as $\lambda \rightarrow
0$;  the high temperature expansion at fixed $\nu$, for $\tilde {\bm r}_{\bm q}$  $\equiv$ $\bm
e_a\epsilon^{ab}q_b\ell_B^2$, is
\begin{equation}
\frac{s(\bm q) - s_{\infty}}{(s_{\infty})^2} = -\left (\frac{ \tilde v(\bm
  q) + \xi v(\tilde {\bm r}_{\bm q} )}{k_BT}\right ) +
 O\left (\frac{1}{T^2}\right ).
\label{ht}
\end{equation}

The correlation energy per flux quantum is given by
\begin{equation}
\bar \varepsilon = \int \frac{d^2\bm q \ell_B^2}{4\pi}
\tilde v(\bm q) \left ( s(\bm q) - s_{\infty}\right ).
\end{equation}
The fundamental duality of the structure function (already apparent in
(\ref{ht}), and derived below) is
\begin{equation}
s(\bm q) - s_{\infty} = \xi \int \frac{d^2\bm q'\ell_B^2}{2\pi}
e^{i\bm q\times \bm q'\ell_B^2} \left ( s(\bm q') - s_{\infty}\right ).
\label{dual}
\end{equation}
This is valid for a structure function calculated using \textit{any}
translationally-invariant density-matrix, and assumes that no additional
degrees of freedom (\textit{e.g.}, spin, valley,  or layer indices)  
distinguish the particles.

Consider the equilibrium state of a system with temperature $T$ and
filling factor $\nu$ with the Hamiltonian (\ref{ham}).  The free
energy  of this state is formally given by $F[\hat \rho_{\rm eq}]$,
where
$\hat \rho_{\rm eq} (T,\nu)$ is the equilibrium density-matrix $Z^{-1}\exp( -
H/k_BT)$
and $F[\hat \rho]$ is the functional
\begin{equation}
F[\hat \rho] = {\rm Tr}\left (\hat \rho (H + k_BT \log \hat \rho)\right),
 \end{equation}
which, for fixed $\nu$,  is minimized when $\hat \rho$ = $\hat
\rho_{\rm eq}$. 
The  APD corresponding to a shear is  $R^a$ $\rightarrow $
$R^a +  \epsilon^{ab}\gamma_{bc}R^c$, parametrized by a symmetric
tensor
 $\gamma_{ab}$ = $\gamma_{ba}$.  Let $\hat \rho (\gamma)$ =
$U(\gamma)\hat \rho_{\rm eq}U(\gamma)^{-1}$, where $U(\gamma)$ is the
unitary operator that implements the APD, and $F(\gamma)$ $\equiv$
$F[\hat \rho (\gamma)]$ = $F[\hat\rho_{\rm eq}]$ +  $O(\gamma^2) $, which is
 is minimized when
$\gamma_{ab}$ = 0.   The free energy per flux quantum  has the expansion
\begin{equation}
f(\gamma) = f(T,\nu) +
{\textstyle\frac{1}{2}}G^{abcd}(T,\nu)\gamma_{ab}\gamma_{cd} +
  O(\gamma^3),
\end{equation}
where $G^{abcd}$  = $G^{bacd}$ = $G^{cdab}$.   
The ``\textit{guiding-center shear modulus}'' (per flux quantum) of
the state
is given by  $G^{ac}_{bd}$  = $\epsilon_{be}\epsilon_{df}G^{aecf}$,
with $G^{ac}_{bd}$ = $G^{ca}_{db}$, and
$G^{ac}_{bc}$ = 0. (Note that in a spatially-covariant
formalism, both stress $\sigma^a_b$ (the momentum current)  and strain
$\partial_cu^d$ (the gradient of the displacement field)  are
mixed-index tensors that are linearly related by the elastic modulus
tensor $G^{ac}_{bd}$.) 
The entropy is left
invariant by the APD, and the only affected  term in the free energy 
is the correlation energy, which can be
evaluated in terms of the deformed structure factor
$s(\bm q,\gamma) $, given by
\begin{equation}
s_{\infty} + \xi
\int \frac{d^2\bm q'\ell_B^2}{2\pi}e^{i\bm q\times \bm q'\ell_B^2} 
(s(\bm q')-s_{\infty}) 
e^{i\gamma^{ab}q_aq'_b\ell_B^2},
\end{equation}
with $\gamma^{ab}$ $\equiv$ $\epsilon^{ac}\epsilon^{bd}\gamma_{cd}$.
This gives $G^{abcd}(T,\nu)\gamma_{ab}\gamma_{cd}$ as
\begin{equation}
\gamma_{ab}\gamma_{cd}\epsilon^{ae}\epsilon^{cf}\int\frac{d^2q\ell_B^2}{4\pi}
\tilde v(\bm q)
q_eq_fs^{bd}(\bm q;T,\nu). 
\end{equation}

Assuming only that the ground state  $|\Psi_0\rangle $ of (\ref{ham}) has  translational
invariance, plus inversion symmetry (so it has vanishing electric
dipole moment parallel to the 2D surface), GMP\cite{GMP} used the SMA
variational
state
$|\Psi(\bm q)\rangle $ $\propto$ $\rho(\bm q)|\Psi_0\rangle$
to obtain an upper bound  $E(\bm q)$ $\le$ $f(\bm q)/s(\bm q)$
to the energy of an excitation  with momentum
$\hbar \bm q$ (or electric dipole moment  $e
\bm e_a \epsilon^{ab}q_b\ell_B^2$), 
where
\begin{eqnarray}
f(\bm q) &=& \int \frac{d^2\bm q'\ell_B^2}{4\pi}
\tilde v(\bm q') \left (2\sin {\textstyle\frac{1}{2}}\bm q\times \bm
    q'\ell_B^2\right )^2
s(\bm q',\bm q),\nonumber  \\
s(\bm q',\bm q) &\equiv& {\textstyle\frac{1}{2}}\left (
s(\bm q' + \bm q) + s(\bm q' - \bm q) - 2s(\bm
    q')\right ).
\end{eqnarray}
Other than noting it was quartic in the small-$q$ limit,
GMP did not not offer any further interpretation of $f(\bm q)$.   It
can now be seen to have the long-wavelength behavior
\begin{equation}
\lim_{\lambda\rightarrow 0}f(\lambda \bm q) 
\rightarrow {\textstyle
\frac{1}{2}} \lambda^4 G^{abcd}q_aq_bq_cq_d\ell_B^4,
\end{equation}
and is controlled by the guiding-center shear-modulus.
Then the SMA result is, at long wavelengths,  at $T=0$,
\begin{equation}
E(\bm q) s(\bm q) \le {\textstyle\frac{1}{2}}
G^{abcd}q_aq_bq_cq_d\ell_B^2.
\end{equation}
This relation also occurs (as an equality)
in the 
phonon spectrum  of the Wigner crystal of 
guiding centers\cite{BM}: if
$\tilde v(\bm q)$ diverges as $\bm q \rightarrow 0, $ the leading
behavior is
\begin{eqnarray}
E(\bm q) &\rightarrow& \left (  {\textstyle\frac{1}{2}}
  G^{abcd}q_aq_bq_cq_d\ell_B^4
\right )^{1/2} \times
\left (\tilde v(\bm q)\right )^{1/2} , \\
s(\bm q) &
\rightarrow& \left (  {\textstyle\frac{1}{2}}
  G^{abcd}q_aq_bq_cq_d\ell_B^4\right )^{1/2} \times
\left (\tilde v(\bm q)\right )^{-1/2} .
\end{eqnarray}

In the case of the incompressible FQH fluid with a gap for excitations
carrying an electric dipole moment, it appears\cite{yang}
that the SMA may give the exact collective mode dispersion in the
long-wavelength limit,
in which case the FQH structure function, at least for 
one-component FQH fluids such as the Laughlin states, has the limit
(at $T$ = 0)
\begin{equation}
\lim_{\lambda\rightarrow 0}s(\lambda \bm q) \rightarrow 
{\textstyle\frac{1}{2}}\lambda^4(G^{abcd}q_aq_bq_cq_d\ell_B^4)/E(0).
\end{equation}
The $O(\lambda ^4)$ behavior of $s(\lambda \bm q)$ at small $\lambda$
 is the fundamental property of FQH fluids identified by
 GMP.

The derivation above finally completes the picture of FQH  incompressibility
initiated by GMP,  by
linking the existence of the  gap to the shear modulus of the fluid, and essentially to
the geometry of flux attachment to its ``elementary droplet'' or
``composite boson'',
 and the energy cost of area-preserving diffeomorphisms that
distort the shape of  correlations  around guiding centers.
This key geometric ingredient appears to be missing in 
previously-proposed descriptions of
FQH incompressibility (``Ginzburg-Landau-Chern-Simons
theory'' of a composite-boson superfluid\cite{glcs}, filling of ``effective Landau levels'' by
composite fermions\cite{jain}, or ``non-commutative Chern-Simons
theory''\cite{suss}),
none of which  derive directly  from the Hamiltonian
(\ref{ham}) and its  algebra  (\ref{gmpa}). 

In the systems studied to date, the 
collective mode energy $E(\bm q)$ has a ``roton minimum''
at finite $\bm q$ = $\pm \bm
q_{\rm \min}$,
with  $2E(\bm q_{\rm min}) < E(0)$, so the lowest energy (quadrupolar) excitations
at $\bm q = 0$ are either roton pairs, or a two-roton bound state\cite{GMP}, and
the long-wavelength limit of the collective mode is hidden  in the
continuum, and presumably damped by decay into roton pairs.
However, there seems no obvious reason why some choice of the interaction
$\tilde v(\bm q)$ could not expose  $E(0)$ by moving it down below the bottom of the
two-roton continuum.

I now outline the derivation of the duality (\ref{dual}).
It will be  convenient to impose periodic boundary conditions on a
fundamental region of area $2\pi N_{\Phi}\ell_B^2$, so the allowed
values of $\bm q$ define a Bravais lattice in reciprocal space with
unit cell area $2\pi/N_{\Phi}\ell_B^2$.  (Indeed, I initially noticed
the duality in numerical exact-diagonalization results for $s(\bm q)$ in such a geometry.)
If both $\bm q$ and $\bm q'$
are allowed reciprocal vectors,
\begin{equation}
\left ( e^{i\bm q\times \bm q'\ell_B^2}\right )^{N_{\Phi}} = 1.
\end{equation}
Then, for an allowed $\bm q$,
\begin{equation}
\left (e^{i\bm q \cdot \bm R}\right)^{N_{\Phi}}|\alpha\rangle =
\left (\eta_{\bm q}\right )^{N_{\Phi}}|\alpha\rangle,
\end{equation}
where $|\alpha\rangle$ is any of the $N_{\Phi}$ linearly-independent
one-particle states in the Landau level, and $\eta(\bm q)$ is $1$ if
$\frac{1}{2}\bm q$ is on the Bravais lattice, and $-1$ otherwise.

There is a recurrence relation, so that for $\bm q$ and $\bm q'$ on the
Bravais lattice,
\begin{equation}
\rho (\bm q + N_{\Phi}\bm q') = \rho(\bm q)\left (\eta_{\bm q'}e^{i\frac{1}{2}\bm
    q\times\bm q'\ell_B^2}\right )^{N_{\Phi}} = \pm
  \rho(\bm q),
\end{equation}
which means that there are exactly $(N_{\Phi})^2$ linearly-independent generators
$\rho(\bm q)$, one of which is equivalent to $\rho(\bm 0)$, and which
define a ``Brillouin zone'' (BZ).  (The algebra (\ref{gmpa}) is now
that of the generators of $U(N_{\Phi})$.)  The two-guiding-center exchange operator
can be written as
\begin{equation}
P_{ij} = \frac{1}{N_{\Phi}}{\sum_{\bm q'\in \text{BZ}}}' 
e^{i (\bm  q'+\bm q)\cdot \bm R_i}e^{-i(\bm  q'+\bm q)\cdot \bm R_j},
\label{exch}
\end{equation}
where the primed sum is over \textit{any} set of  $(N_{\Phi})^2$ reciprocal
vectors $\{\bm q'\}$  that define a BZ.
The freedom of choice of the BZ means that  $\bm q$  in (\ref{exch}) can be chosen
arbitrarily on the Bravais lattice. 
This expression can be rewritten as
\begin{equation}
P_{ij}e^{i\bm q\cdot \bm R_i}e^{-i\bm q\cdot\bm R_j}
= \frac{1}{N_{\Phi}}{\sum_{\bm q'\in \text{BZ}}}'e^{i\bm q\times \bm q'\ell_B^2} 
e^{i \bm  q'\cdot \bm R_i}e^{-i\bm  q'\cdot \bm R_j},
\end{equation}
Take the expectation value, and sum over $i, j$, noting that $P_{ij}$ = $\xi$ for $i\ne j$,
 $P_{ii}$ = $N_{\Phi}$, and, for
$\rho(\bm q)$ = $\sum_i\exp i \bm q\cdot\bm  R_i$ not equivalent to
$\rho(\bm 0)$, that $\langle \rho(\bm q)\rho(-\bm q)\rangle$ =
$N_{\Phi}s(\bm q)$, and $\langle N^2\rangle - \langle N\rangle^2$ =
$N_{\Phi}s(0)$, $\nu$ = $\langle N\rangle/ N_{\Phi}$. Then
\begin{equation}
s(\bm q)  - s_{\infty} = \frac{\xi}{N_{\Phi}}{\sum_{q'\in
    \text{BZ}}}'e^{i\bm q\times\bm q'\ell_B^2}\left (s(\bm
  q')-s_{\infty}\right ).
\end{equation}
No assumptions other than translational invariance (and full symmetry
or antisymmetry under guiding-center exchange) of the density-matrix appear to
have been made in this derivation.  In particular, no rotational
invariance has been assumed.  Taking the thermodynamic limit
$N_{\Phi}\rightarrow \infty$ leads directly to (\ref{dual}).

Further insight  is provided by the expansion
\begin{equation}
\delta \rho(\lambda \bm q) = \delta N  + i\lambda \epsilon^{ab}q_aK_b\ell_B^2 -
\lambda^2 q_aq_b\Lambda^{ab}\ell_B^2 + O(\lambda^3),
\end{equation}
where $\delta N$ = $N-\langle N\rangle$ commutes with $\delta\rho(\bm q)$, and
\begin{eqnarray}
&& [K_a,K_b] = 0, \quad [\Lambda^{ab},K_c] =
-{\textstyle\frac{i}{2}}\left (\delta^a_c\epsilon^{bd} +
  \delta^b_c\epsilon^{ad}\right )
K_d, \nonumber \\
&&
[\Lambda^{ab},\Lambda^{cd}] = -{\textstyle\frac{i}{2}}\left (
\epsilon^{ac}\Lambda^{bd} + \epsilon^{ad}\Lambda^{bc} +
a\leftrightarrow b\right ).
\end{eqnarray}
Here  $ K_a$ are the generators of translations,
$[K_a,\delta\rho(\bm q)]$ = $q_a\delta\rho(\bm q)$, and
 $e\bm
e_a\epsilon^{ab}K_b\ell_B^2$ is the total electric dipole moment.  They
annihilate an incompressible FQH state: $ K_a |\Psi_0\rangle$ = 0.  
The three independent components of the symmetric tensor
$\Lambda^{ab}$ are the generators of linear  APD's of the
guiding centers,
and obey the Lie algebra of the generators of the group $SL(2,R)$,
isomorphic to $SO(2,1)$, with Casimir $\det \Lambda$\cite{haldanearxiv}.  Then at long wavelengths,
\begin{equation}
s(\lambda \bm q) \rightarrow \lambda^4
 \Gamma_S^{abcd}q_aq_bq_cq_d\ell_B^4,\\
\end{equation}
where
\begin{equation}
\Gamma_S^{abcd} = (N_{\Phi})^{-1}\left (
{\textstyle\frac{1}{2}}\langle \{\Lambda^{ab},\Lambda^{cd}\}\rangle -
\langle \Lambda^{ab}\rangle\langle\Lambda^{cd}\rangle \right
).
\end{equation}
A bound to the tensor $ \Gamma_S^{abcd}$ can be set using the related tensor
\begin{eqnarray}
 \Gamma_A^{abcd} &=& (N_{\Phi})^{-1}
{\textstyle\frac{i}{2}}\langle [\Lambda^{ab},\Lambda^{cd}]\rangle
\nonumber \\
&=&
{\textstyle\frac{1}{2}}\left ( \epsilon^{ac} \Gamma_H^{bd} +
  \epsilon^{ad} \Gamma_H^{bc} + a\leftrightarrow b\right ),\\
 \Gamma^{ab}_H &=&{\textstyle\frac{1}{2}} (N_{\Phi})^{-1}\langle \Lambda^{ab}\rangle.
\end{eqnarray}
Viewed as a $3\times 3$  Hermitian matrix, $M^{(ab),(cd)}$ =
$ \Gamma_S^{abcd}\pm i  \Gamma_A^{abcd}$ is positive (has no negative
eigenvalues). This means $\Gamma^{ab}_H$ can set a lower bound
to $\Gamma_S^{abcd}q_aq_bq_cq_d$\cite{haldanearxiv}.

Note that $\eta_H^{abcd}$  =  $(eB/2\pi) \Gamma_A^{abcd}$ is the
guiding-center
contribution\cite{haldanearxiv} to
the so-called  dissipationless ``Hall viscosity'' tensor (there is also a Landau-orbit contribution\cite{avron}).
Physically, $ \Gamma_A^{abcd}$ gives  the 
local stress (momentum current)  induced by the linear response of the uniform
FQH state to a non-uniform electric field:
\begin{equation}
\sigma^a_e(\bm r) = \frac{e}{2\pi}\epsilon_{eb}\left( \Gamma_A^{abcd} +
  \Gamma_A'^{abcd}\right )\partial_cE_d(\bm r),
\label{xx}
\end{equation}
where $\Gamma_{A}'^{abcd}$ in (\ref{xx}) is the 
Landau-orbit contribution\cite{avron}.

If rotational invariance with a unimodular metric $g_{ab}$ is present,
then guiding-center rotations are generated by $L$ =
$g_{ab}\Lambda^{ab}$, with
$[H, L]$ = 0, and $\langle L^2\rangle -\langle L\rangle^2$ = 0, from which
$ \Gamma_S^{abcd}$ =$\alpha \left ( g^{ac}g^{bd} + g^{bc}g^{ad} - g^{ab}g^{cd}\right
)$, and $s(\bm q)$  $\rightarrow$
$\alpha(g^{ab}q_aq_b\ell_B^2)^2$
at long wavelengths.
Also $\Gamma_H^{ab}$ = $\frac{1}{4}\bar \ell g^{ab}$, where $\bar \ell
$ = $\langle L\rangle/N_{\Phi}$, giving the bound $\alpha \ge
\frac{1}{4}|\bar \ell |$.

Incompressible FQHE states may be partly classified by  their ``elementary  droplet''
or ``composite boson'' that consists of $p$ particles in
$q$ orbitals (``flux attachment'' $q$), where for $\nu$ = $p/q$, $p$ and $q$ are the
smallest
integers obeying the statistical selection rule $(-1)^{pq}$ = $\xi^p$
which ensures that pairs of droplets behave as bosons when
exchanged.     The elementary droplet also has a topologically-quantized\cite{haldane2011}  (integer
or half-integer) ``guiding-center spin''   $\bar s$ (a 2D orbital spin unrelated
to the Pauli spin of the electron), and $\bar \ell $ = $\bar s/q$,
with $|\bar s|  < \frac{1}{2}pq$.
If there is rotational invariance, and $\bar N$ elementary droplets
are combined to form a circular droplet of  FQH fluid centered at the
origin,   with no internal excitations, the total guiding-center angular momentum is
\begin{equation}
 L_0 = {\textstyle\frac{1}{2}}pq\bar N^2 + \bar s \bar N, \quad
\frac{g_{ab}}{2\ell_B^2}\sum_iR_i^aR_i^b|\Psi_0\rangle =
L_0|\Psi_0\rangle.
\end{equation}
The first term in $L_0$ is the uniform-disk contribution
$\frac{1}{2}\nu (N_{\Phi})^2$, $N_{\Phi}$ = $q\bar N$; the second term is $\bar \ell N_{\phi}$\cite{read}.

The bound $\alpha \ge  |\bar s|/4q$ is satisfied as an equality\cite{GMP,haldanearxiv} in the
case of certain polynomial model wavefunctions (Laughlin, Moore-Read,
\textit{etc.})
that can be viewed as correlators of chiral conformal field theories,
and have rotational invariance.   Read and Rezayi\cite{RR} have
argued not only that the bound
satisfied by these model states will remain satisfied as an equality in the presence of
perturbations that maintain rotational invariance, but  also claim that this
must be
be generically true for ``all lowest-Landau level'' FQH states with
rotational invariance.
The counter-propagating
principal hierarchy/Jain sequence  $\nu$ = $n/(4n-1)$, $n$ =
$1,2,\ldots $, descending from the $\nu = \frac{1}{3}$ Laughlin
state, with $p$ = $n$, $q$ = $4n-1$, has $\bar s$ = $\frac{1}{2}n(n-3)$.
If the claim  of Ref.\cite{RR} were generally correct, the $\nu$ =
$\frac{3}{11}$ state would have vanishing $\alpha$. Instead, satisfaction of the
bound as an equality
is likely to occur  only in \textit{maximally-chiral} FQH states without
counter-propagating components (in this sequence, $n$ = 1 only). 
 The property of maximal chirality
of a FQH state
can in principle be identified from  its entanglement
spectrum\cite{Li},
without reference to model wavefunctions.

In summary, I have exhibited a self-duality of the guiding-center
structure function $s(\bm q)$ of translationally-invariant states of a
partially-filled Landau level (a ``quantum geometry'' problem) and
used it to interpret the long-wavelength collective excitations of FQH states
in terms of a shear modulus. Bounds on the $O(q^4)$ behavior of $s(\bm
q)$ were also obtained.

This work was supported in part  by DOE grant {DE}-{SC0002140}. 

\end{document}